\documentstyle[aps, twocolumn, epsfig]{revtex}

\narrowtext
\begin{document}

\input epsf.sty
\newcommand{\infig}[2]{\begin{center}\mbox{ \epsfxsize #1
                       \epsfbox{#2}}\end{center}}

\draft 
\wideabs{
\title{An optical kaleidoscope using a single atom}
\author{P.\ Horak and H.\ Ritsch}
\address{Institut f\"{u}r Theoretische Physik, Universit\"{a}t Innsbruck,
Technikerstr. 25, A-6020 Innsbruck,  Austria}
\author{T.\ Fischer, P.\ Maunz, T.\ Puppe, P.W.H.\ Pinkse, and G.\ Rempe}
\address{Max-Planck-Institut f\"{u}r Quantenoptik, Hans-Kopfermann-Str. 1,
D-85748 Garching, Germany}
\date{\today}
\maketitle

\begin{abstract}
A new method to track the motion of a single particle in the field of a
high-finesse optical resonator is described. It exploits near-degenerate
higher-order Gaussian cavity modes, whose symmetry is broken by the phase shift
on the light induced by the particle. Observation of the spatial intensity
distribution behind the cavity allows direct determination of the particle's
position with approximately wavelength accuracy. This is demonstrated by
generating a realistic atomic trajectory using a semiclassical simulation
including friction and diffusion and comparing it to the reconstructed path.
The path reconstruction itself requires no knowledge about the forces on the
particle.
\end{abstract}
%\pacs{PACS numbers: 42.50.Ct, 32.80.Pj, 42.50.Vk}
}

\narrowtext
%\onecolumn

\renewcommand{\vec}[1]{{\bf {#1}}}
\newcommand{\vecr}{\vec{r}}
\newcommand{\vecp}{\vec{p}}
\newcommand{\waist}{{\mathrm w_0}}
\newcommand{\upm}{u_{pm}}
\newcommand{\upn}{u_{pn}}
\newcommand{\gpm}{g_{pm}}
\newcommand{\apm}{\alpha_{pm}}
\newcommand{\apn}{\alpha_{pn}}

%\section{Introduction}
In a variety of pioneering experiments in the past few
years\cite{MabuchiOL96,MunstermannOptCom99,ChildsPRL96} it has been
demonstrated and widely exploited that a single near-resonant atom can
significantly influence the field dynamics in a microscopic high-finesse
optical resonator. Vice versa, the light field also influences the motion of a
cold atom, which leads to an intricate dynamical interplay of atomic motion and
field dynamics~\cite{HoodPRL98,MunstermannPRL99}. As a striking example,
trapping of a single atom in the field of a single photon has become
feasible\cite{RempeNature,KimbleScience}. This was experimentally substantiated
by analyzing the characteristics of the measured output field. The time
variation of the transmitted intensity shows very good agreement with
theoretical simulations\cite{DohertyPRA97} of the confined three-dimensional
motion of the atom in the cavity light field including friction and
diffusion~\cite{HorakPRL97,PinkseJMO00,GanglEPD00}. Carrying this analysis
further it was even possible to associate piecewise reconstructed trajectories
with recorded time-dependent intensity curves\cite{KimbleScience,DohertyPRA00},
utilizing the knowledge of the near-conservative potential. However, the
reconstruction was possible only for atoms with sufficiently large and
conserved angular momentum around the cavity axis and can only be done up to an
overall angle and the direction of rotation. The reason was that only a single
cavity mode, the TEM${}_{00}$ mode, was used. Consequently, only a single
spatial degree of freedom of the atom could be extracted directly from a
measurement of the transmitted field.

In this letter, we propose a new method to obtain two-dimensional information
on the atom's position. We consider higher-order frequency-degenerate
transversal modes. Typical examples are the Hermite-Gaussian (HG) or the
Laguerre-Gaussian (LG) modes, which show in the transverse plane a rectangular
matrix of intensity minima and maxima or a pattern of concentric rings,
respectively. In such a configuration, the atom can redistribute photons from
one mode to the other. Moreover, it induces mode frequency shifts and losses
strongly dependent on the atomic position. In particular, the symmetry of the
intracavity field determined by the cavity and pump geometry will be perturbed
and characteristic optical patterns containing information on the atomic
position appear. The effect reminds of a toy-kaleidoscope in which small
objects in a symmetric arrangement of mirrors create beautiful patterns. Our
technique yields much more information on the atomic position and motion as
compared to the single-mode case, and allows to extract the atomic position
directly from a field measurement.

%%Model
To treat this problem quantitatively we generalize previous semiclassical
models of dynamical cavity QED in the strong-coupling regime to include finite
sets of nearly degenerate eigenmodes. For a weakly saturated atom we derive a
coupled set of equations for the mode amplitudes and the atomic center-of-mass
motion. To be specific, let us consider a single two--level atom with
transition frequency $\omega_a$ and line width $\Gamma$ (half width at half
maximum) moving inside a high-finesse optical resonator with transversal LG
eigenmodes $u_{pm}(\vecr)$, where $p$ is the radial mode index and $m$ is the
azimuthal mode index~\cite{modefunctionref},
\begin{eqnarray}%
\upm(\rho,\theta,z) &=&C_{pm} \cos(kz)
   e^{-\frac{\rho^2}{\waist^2}+i m \theta} \nonumber\\
   & & \times   (-1)^p (\frac{\rho\sqrt{2}}{\waist})^{|m|}
                L_p^{|m|}(\frac{\sqrt{2}\rho^2}{\waist^2}),
\end{eqnarray}%
where $L_n^\alpha$ is the generalized Laguerre polynomial.
% These modes are degenerate for a given $n_0 = 2p+|m|$.
The normalisation parameters, $C_{pm}$, are chosen such that $\int
|\upm(\rho,\theta,z)|^2 {\rm d}V = d\waist^2\pi/4=V_{00}$ (the TEM${}_{00}$
mode volume), where $\waist$ is the cavity waist and $d$ the cavity length. At
each spatial point the local atom-mode couplings are
$\gpm(\vecr)=g_0\upm(\vecr)$, where $g_0$ is the maximum coupling of the
TEM${}_{00}$ mode.
For simplicity, we assume that the mirrors are ideally spherical and have a
uniform coating. Then, all these modes have a common eigenfrequency
$\omega$ and field decay rate $\kappa$. However, the model can be extended to
incorporate non-degenerate modes in a straightforward manner. The cavity is
also assumed short compared to the Rayleigh length of the mode so that the
wavefronts are approximately plane with $z$-dependence $\cos(2\pi z/\lambda)$.
The resonator is externally driven by a coherent pump field of frequency
$\omega_p$ which pumps the modes with strengths $\eta_{pm}$. Assuming low
atomic saturation, we can adiabatically eliminate the internal atomic dynamics
and treat the atom as a linearly polarizable particle, which induces a
spatially dependent phase shift and loss. In the semiclassical
limit\cite{semipaper}, where we consider the center-of-mass motion of the atom
classically, we can derive the following set of coupled differential equations
for the mode amplitudes $\alpha_k(t)$, the atomic position $\vecr_a(t)$ and
momentum $\vecp_a(t)$:
\begin{eqnarray}
\dot{\vecr}_a & = & \frac{\vecp_a}{M}, \nonumber \\
\dot{\vecp}_a & = & -U_0\sum_{m,n} \nabla(\upm \upn^*)
   \apm \apn^*\nonumber \\
 & & +i\gamma\sum_{m,n}
   (\upm\nabla u^*_{pn} - u^*_{pn}\nabla \upm)
   \apm \alpha^*_{pn} + \chi, \nonumber \\
\dot{\alpha}_{pm} & = & \eta_{pm} + (i\Delta-\kappa)\apm\nonumber \\
 & & -(iU_0+\gamma)\upm\sum_n u^*_{pn} \apn +
   \xi_{pm}.
\label{modedyn}
\end{eqnarray}
Here, $M$ is the atomic mass, $U_0 = g_0^2/(\omega_p-\omega_a)$ is the
single-photon optical light shift, $\gamma = \Gamma
g_0^2/(\omega_p-\omega_a)^2$ the spontaneous emission rate for a single-photon
field, $\Delta = \omega_p-\omega$ the cavity-pump detuning, and $\chi$ and
$\xi_{pm}$ are Gaussian random variables which model momentum and cavity field
fluctuations, respectively.

%% Steady state field
As an example, we consider an atom at rest at a given position $\vecr_a$
coupled to the three degenerate cavity modes with $(p,m)=(1,0),(0,-2),(0,2)$.
It is then possible to solve equations (\ref{modedyn}) for the mean stationary
field amplitudes $\apm^{\mbox{\tiny stat}}(\vecr_a)$, which in a parametric way
depend on the atomic position. We get:
\begin{equation}
    \alpha_{pm}^{\mbox{\tiny stat}}=\frac{\eta_{pm}^*}{i\Delta-\kappa}
   +\frac{iU_0+\gamma}{i\Delta-\kappa}\upm^*(\vecr_a){\cal E}_0(\vecr_a),
\end{equation}
where ${\cal E}_0$ is the electric field at the position of the atom,
\begin{equation}
{\cal E}_0(\vecr_a)  =  \frac{\sum_{p,m} \upm(\vecr_a)\eta_{pm}^*}{
   (i\Delta-\kappa)-(iU_0+\gamma)\sum_{p,m}|\upm(\vecr_a)|^2}.
\end{equation}
Note that the set of mode amplitudes $\upm(\vecr_a)$ at the position of the
atom may be written as a vector, which can then be rotated into the form
$(u_{\rm eff}(\vecr_a),0,0)$. Since any linear combination of the modes can be
considered as a mode as well, the atom-field dynamics for an atom at rest thus
reduces to the case of a single mode $u_{\rm eff}$. This allows to derive
simple analytical expressions for the steady state. For a moving atom, this
approach must be generalized, but still helps to find analytical expressions
for friction and diffusion coefficients for the atomic motion
\cite{FischerNJP}.

%Figure 1

Figure\ 1 shows the steady state field intensities for the empty cavity and for
two different atomic positions. For the chosen parameters (rubidium atoms,
$(g_0,\Gamma,\kappa)=2\pi\times (16,3,1.5)\,$MHz,
$(\eta_{10},\eta_{0-2},\eta_{02}) = 2\pi\times (6.4,0,0)\,$MHz,
$\Delta=-2\pi\times 2.25\,$MHz and $\omega_p-\omega_a=-2\pi\times 114\,$MHz,
leading to $U_0=\Delta$, $\gamma=2\pi\times 60\,$kHz) the atom distributes
photons between the cavity modes and changes their relative phases in such a
way that a local maximum of the total field intensity is created near the
position of the atom. By a change of the detuning between the pump laser, the
cavity modes, and the atomic transition, a local minimum can also be achieved.
Figure 1 shows the effect of the atom position on the shape and overall
intensity of the stationary field distribution.

This dependence of the cavity field on the atomic position suggests that
measuring the cavity output field distribution yields ample information on the
atomic motion. In fact, it can be shown that the functions
$\alpha_{pm}^{\mbox{\tiny stat}}(\vecr_a)$ can be inverted almost everywhere to
yield the atomic position in three dimensions. Of course one is limited by the
common symmetries of all modes. For instance, for the system considered in
Fig.~1, a $180^\circ$ rotation around the cavity axis forms a symmetry
operation. Hence, the reconstruction of the atomic position from the cavity
field will always yield two equivalent positions. Other symmetry operations are
a shift of $\lambda/2$ in direction of the cavity axis and a reflection at the
nodes or antinodes of the standing wave. Another limitation is that an atom
cannot be detected close to the (transversal) nodes of the pumped mode (see
Fig. 1a). However, it is possible to determine directly from the photodetector
signals whether the atomic position can be obtained or not: reconstruction is
possible if the transmission signal with an atom differs from the signal of an
empty cavity. Also, the nodal areas are small and one is free to alternate
rapidly between different pump geometries. Alternatively, one can change the
pump geometry online when the atom approaches the nodal area of the pumped
mode.

Although in principle a full three-dimensional atomic trajectory can be
reconstructed, the method encounters some complications. The
longitudinal motion is in general too fast to be resolved experimentally. This
amounts to replacing the coupling constant $g_0^2$ by its longitudinally
averaged value. A two-dimensional reconstruction still works in this case, even
if the precise factor by which the coupling is reduced is unknown. Second, a
single atom must redistribute enough photons among the cavity modes. This
requires values of $U_0$ of the order of the cavity decay rate $\kappa$ and
hence the strong-coupling regime of cavity QED. Third, our arguments above are
based on a stationary cavity field. For an atom moving in the $xy$ plane, we
thus have to assume that the cavity field follows the atomic motion
adiabatically. This implies slow atomic motion and not too large optical forces
in this plane, which together with the constraint of strong coupling is
tantamount to relatively low intracavity photon numbers. Fourth, in an actual
experiment the exact intracavity photon number can only be deduced from the
number of photons emitted by the cavity. The measured cavity output is subject
to statistical fluctuations (shot noise), which become significant in the few
photon limit. This limits the accuracy to which the cavity field, and therefore
the atomic position, can be known.

In the following we will demonstrate with a sample trajectory that all of these
conditions can be met simultaneously and that the numerical reconstruction of
the atomic path is possible using existing high-finesse optical resonators
\cite{RempeNature,KimbleScience}. For simplicity we will assume a quasi
two-dimensional situation where the atom is trapped in the longitudinal cavity
direction at an antinode ($z=0$) of the standing wave during the full
interaction time.

%% Semiclassical simulation of atomic dynamics and field amplitudes

In a first step we create a sample trajectory for a single atom traversing the
resonator by integrating the stochastic equations of motion (\ref{modedyn}) for
given initial atomic position and velocity. This procedure includes all
reactive and dissipative optical forces which the cavity field imposes on the
atom~\cite{semipaper}, the back-action of the atom on the cavity field, as well
as the momentum and cavity field diffusion. A resulting trajectory is depicted
by the solid curve in Fig.\ 2. The atom enters the resonator from below. By
chance, the atom encircles the cavity axis a few times before it is ejected
again.

%Figure 2
%Figure 3

%% Reconstruction of the field amplitude

The generated trajectory allows us to simulate a realistic cavity output
signal. We assume an arrangement of 16 photodetectors at the cavity output port
each counting the numbers of photons detected in equally sized sectors
extending from the axis to the edges of the cavity and covering an angle of
22.5 degrees. Due to the symmetry of the system, the signals from opposing
detectors can be added without loss of information. To reduce the shot noise,
we integrate the measured photon flux at each detector over a time interval of
100 cavity decay times $1/\kappa$. In Fig.\ 3 we plot the simulated photon flux
for two out of the eight pairs of photodetectors.

We can use the generated time-dependent field pattern to numerically
reconstruct the particular atomic trajectory. This involves several steps.
First, for each time step we determine the atomic position by a least-square
comparison with a list of precalculated detector outputs corresponding to the
steady-state field distribution $\alpha_{pm}^{\mbox{\tiny stat}}(\vecr_a)$ of
any atomic position on a discrete spatial grid. Because of the twofold spatial
symmetry, we always obtain two equivalent points in that way. Hence, from a
chosen initial point we have selected the one point out of these two that forms
a continuous curve as a function of time. The spatial points obtained in that
way are indicated by the crosses in Fig.\ 2. The initial position of the atom
is usually known in experiments where laser-cooled atoms are injected into the
cavity. For the pump geometry chosen here, there is a dark ring where the atom
does not couple to the pumped mode. Close to this ring, the reconstruction is
difficult. As mentioned before, there are ways around this problem, but here
the corresponding crosses are simply left out.

According to the temporal noise of the measured photon fluxes (Fig.\ 3), the
reconstructed atomic positions show a certain spatial spread. Since for the
given parameters the stochastic forces in Eqs.~(\ref{modedyn}) are much smaller
than the Hamiltonian forces, we finally fit a smooth curve to this discrete set
of data. The resulting reconstructed trajectory is shown by the dashed curve in
Fig.\ 2. Note that the curve rotated by $180^\circ$ forms a completely
equivalent solution to the problem. Comparing the reconstructed with the
original trajectory we note that for the depicted area close to the cavity axis
the reconstruction works well with an accuracy which is of the order of an
optical wavelength.

%% Real-life considerations
The proposed detector arrangement might be constructed by segment mirrors
imaging onto an array of single-photon counting detectors. Probably more
practically is direct imaging on a high-sensitive high-speed camera. The
construction of a suited cavity will be challenging. Scatter, misalignment and
deformation of the high-reflectivity mirrors must be kept to a minimum to
prevent breaking of the cylindrical symmetry, which could lift the frequency
degeneracy of the modes by a too large amount.

%% Ring-cavity similarity
It is interesting to note that positive or negative values for $m$ in the
Laguerre-Gaussian modes correspond to actual left- and right-handed rotation of
the light around the cavity axis. Hence, there is a striking similarity between
Fabry-Perot cavities sustaining higher-order Laguerre-Gaussian modes and ring
cavities. Similar dynamics as predicted in Ref.~\cite{GanglPRA00} can therefore
be expected in the system proposed here.

%% Confocal resonator
An extension of the idea presented here is to use a cavity where modes with
different longitudinal mode index are degenerate. %\cite{ConfocalModeCoupling}.
In that case, one can choose a combination of modes with opposite parity to
break the $180^\circ$ rotation symmetry. An example is a LG mode with even $m$
degenerate with another one with odd $m$. In the extreme case of a confocal
cavity, this will lead to a single field maximum near the position of the atom.

%Conclusions
In conclusion, we have shown how a high-finesse microcavity can be used as a
real-time single-particle detector with high spatial resolution. In contrast to
conventional single-atom detection schemes, the cavity works as a
phase-contrast microscope enhanced by the inherent multi-path interference of
the high-finesse cavity. The method does not rely on fluorescence and, hence,
can also work for particles without a closed optical transition. In contrast to
the single-mode case, the information in the field pattern of sufficiently
high-order modes should also allow one to spatially resolve several particles.
It can be expected that the scheme presented here will first be demonstrated
with single atoms in presently available high-finesse cavities, but
applications extend beyond this system. The method should, in principle, also
be applicable to large (bio)molecules in vacuum or even in pure watery
solution.

P.~H.\ and H.~R.\ acknowledge support by the Austrian Science Foundation FWF
(project P13435-TPH).

\begin{figure}
\infig{5cm}{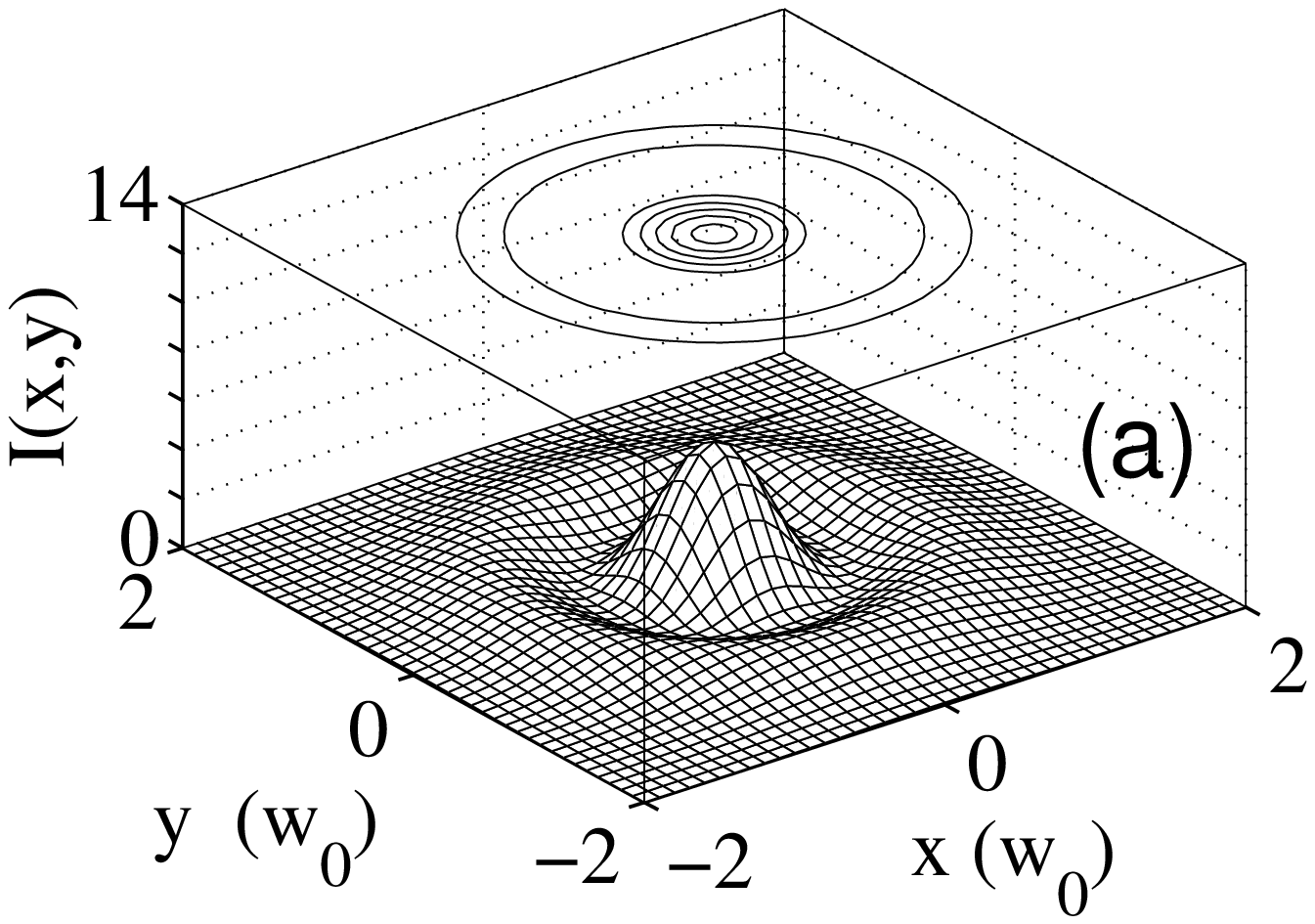}
\infig{5cm}{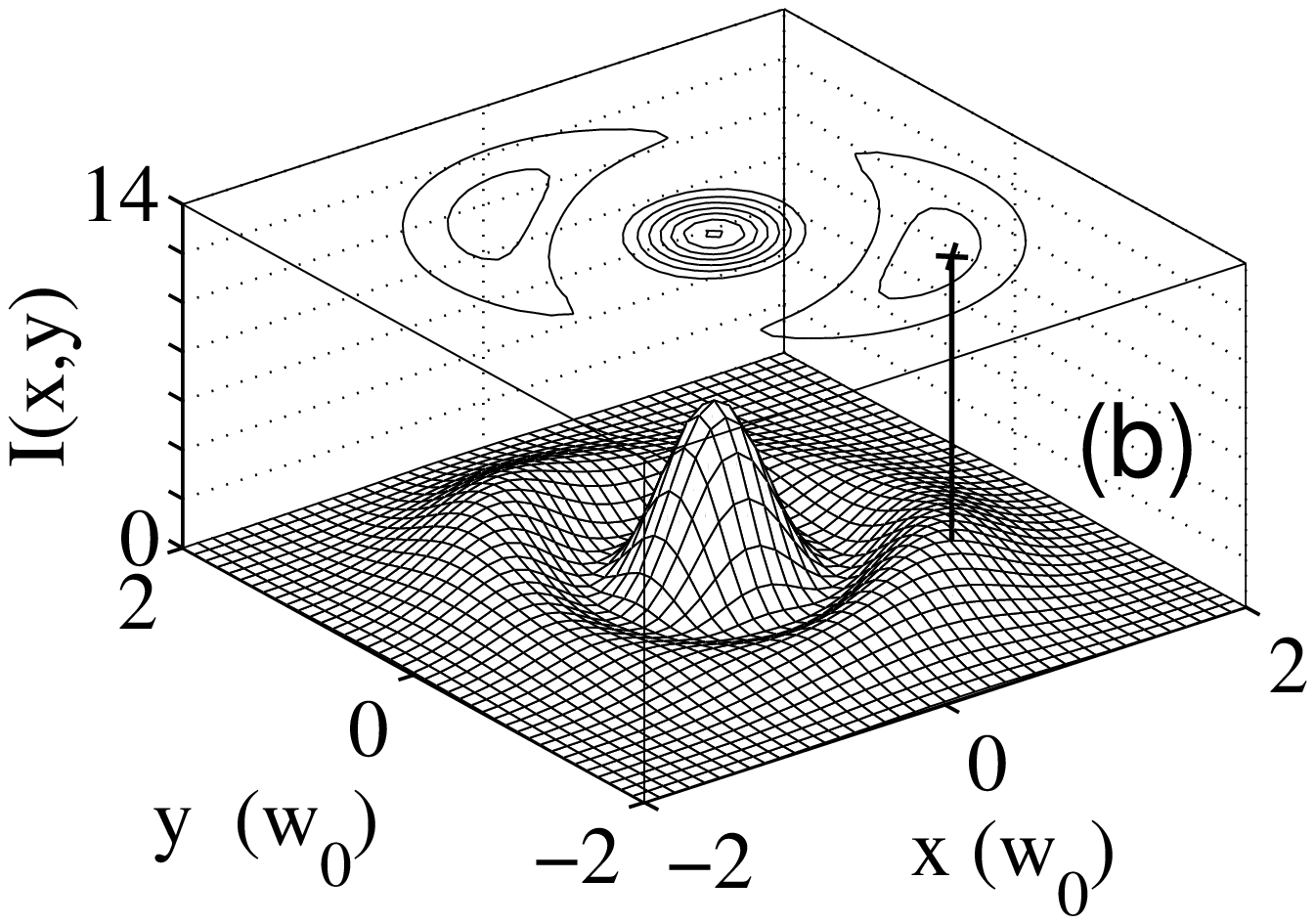}
\infig{5cm}{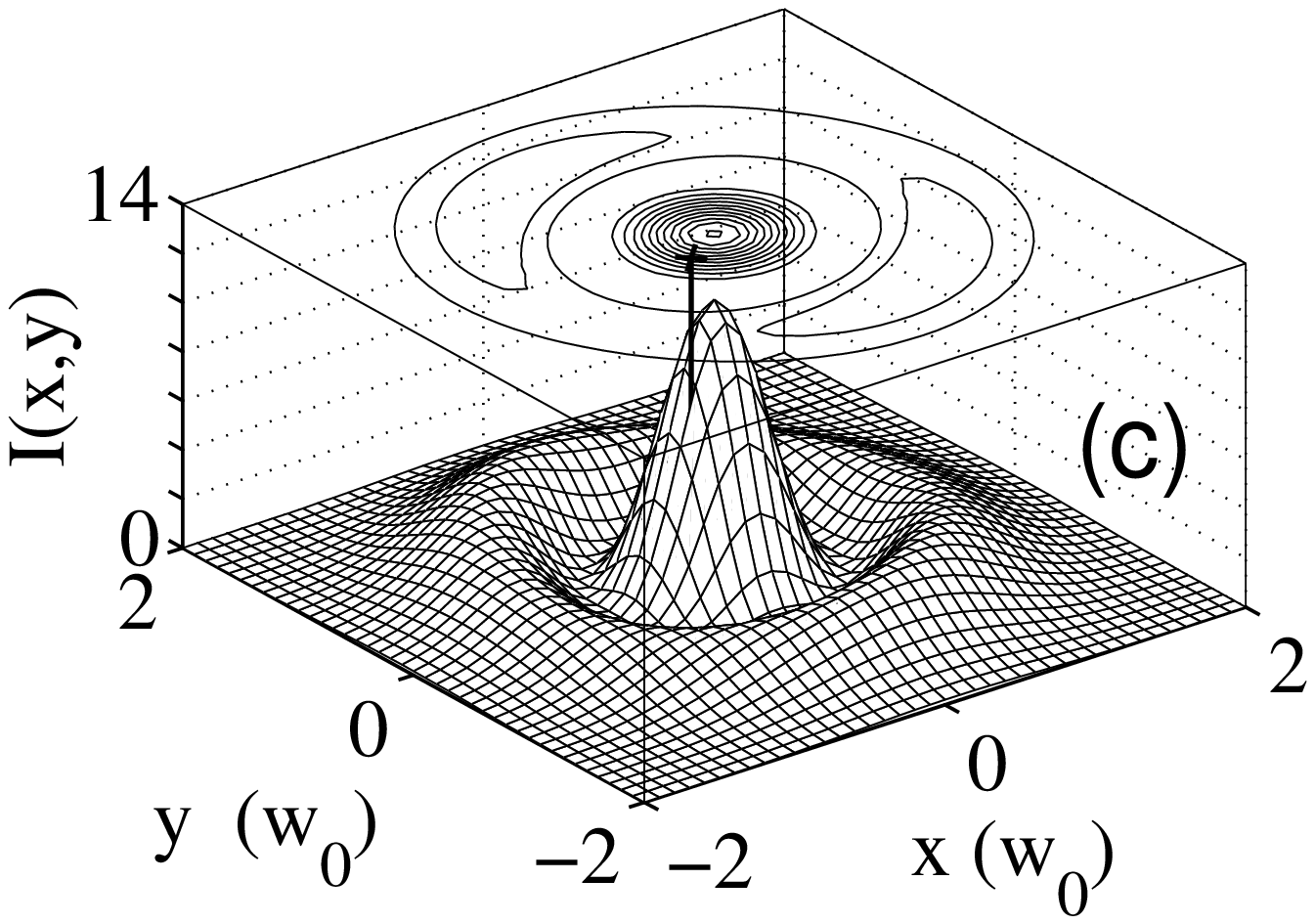} 
\caption{Transversal spatial intensity pattern (arbitrary units) of
the stationary cavity field for the empty cavity (a) and two atomic positions
indicated by the thick vertical line and a cross (b), (c).}
\end{figure}

\begin{figure}
\infig{6cm}{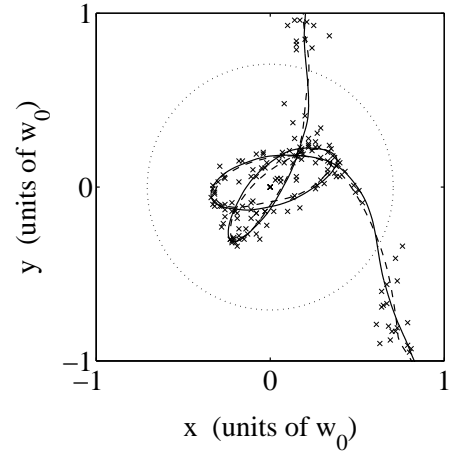}
\caption{Central part of Fig.\ 1(a) with the
simulated atomic trajectory (solid curve), reconstructed atomic positions
(crosses), and fitted atomic path (dashed curve). The atom enters with a
velocity of $12\,$cm/s and the total trajectory takes
$~5300\,\kappa^{-1}=0.56\,$ms. The dashed circle indicates the dark ring of the
pumped cavity mode $u_{10}$. The waist of the TEM$_{00}$ mode is
$\waist=29\,\mu$m.}
\end{figure}

\begin{figure}
\infig{6cm}{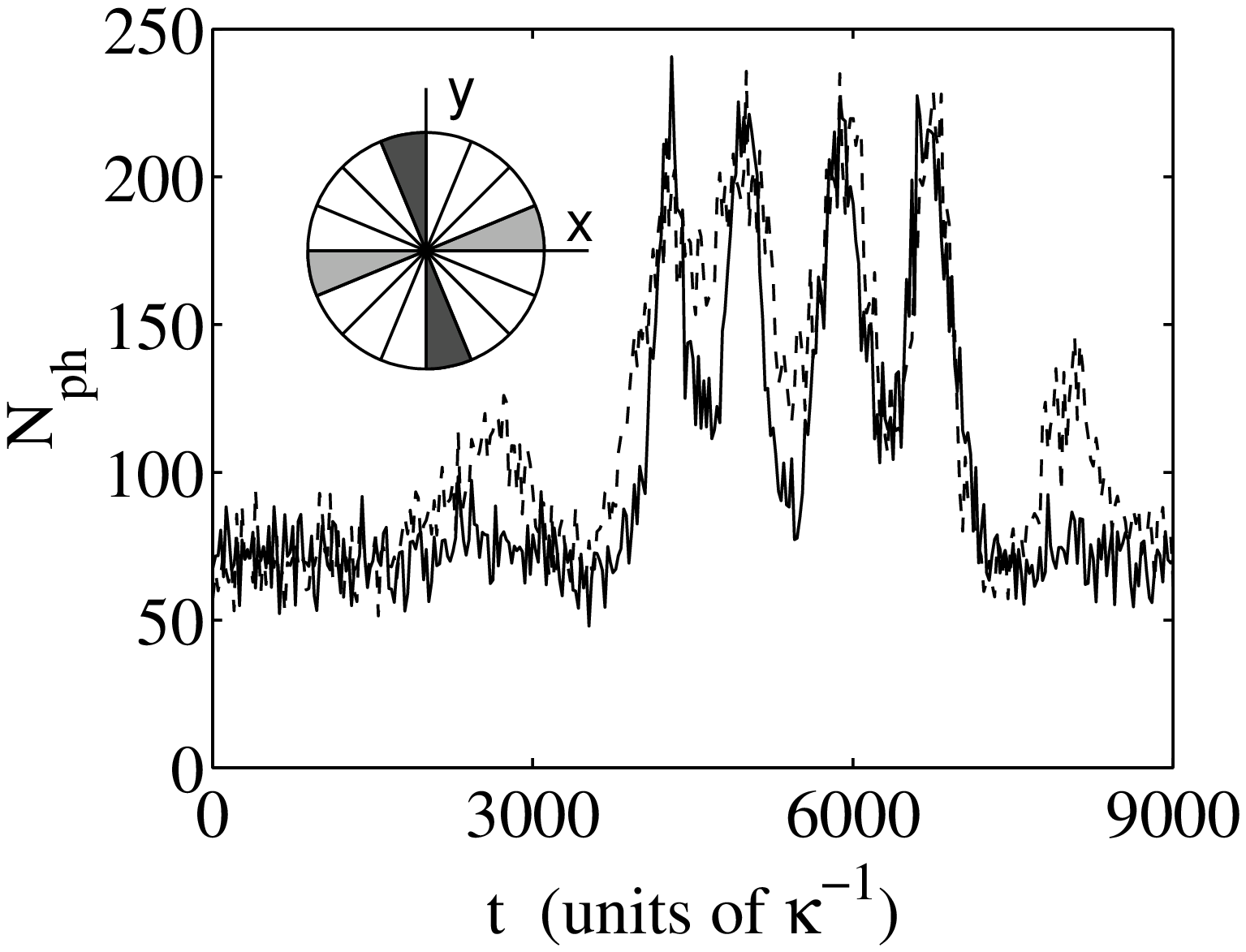} 
\caption{Simulated photon flux (number of photons
$N_{\rm ph}$ per 100 $\kappa^{-1}$) measured at two out of 8 photodetector
pairs as a function of time (units of $\kappa^{-1}$). The inset depicts the
arrangement of the detectors on the cavity output, shaded areas correspond to
the plotted curves.}
\end{figure}


\begin{references}

\bibitem{MabuchiOL96}
H. Mabuchi, Q.A. Turchette, M.S. Chapman, H.J. and Kimble, 1996, Opt.~Lett.
{\bf 21}, 1393 (1996).

\bibitem{MunstermannOptCom99}
P.~M\"unstermann, T.~Fischer, P.W.H.~Pinkse, and G.~Rempe, Opt.~Commun. {\bf
159}, 63 (1999).

\bibitem{ChildsPRL96}
J.J. Childs, K. An, M.S. Otteson, R.R. Dasari, and M.S. Feld, Phys. Rev. Lett.
{\bf 77}, 2901 (1996).

\bibitem{HoodPRL98}
C.J.~Hood, M.S.~Chapman, T.W.~Lynn, and H.J.~Kimble, Phys.~Rev.~Lett.~{\bf 80},
4157 (1998).

\bibitem{MunstermannPRL99}
P.~M\"unstermann, T.~Fischer, P.~Maunz, P.W.H.~Pinkse, and G.~Rempe,
Phys.~Rev.~Lett.~{\bf 82}, 3791 (1999).

\bibitem{RempeNature}
P.W.H.~Pinkse, T.~Fischer, P.~Maunz, and G.~Rempe,
Nature {\bf 404}, 365 (2000).

\bibitem{KimbleScience}
C.J.~Hood, T.W.~Lynn, A.C.~Doherty, A.S.~Parkins, and H.J.~Kimble, Science {\bf
287}, 1447 (2000).

\bibitem{DohertyPRA97}
A.C. Doherty, A.S. Parkins, S.M. Tan, and D.F. Walls, Phys. Rev. A {\bf 56},
833 (1997).

\bibitem{HorakPRL97}
P.~Horak, G.~Hechenblaikner, K.M.~Gheri, H.~Stecher, and H.~Ritsch,
Phys.~Rev.~Lett.~{\bf 79}, 4974 (1997).

\bibitem{PinkseJMO00}
P.W.H. Pinkse, T. Fischer, P. Maunz, T. Puppe, and G. Rempe, J. Mod. Opt. {\bf
47}, 2769 (2000).

\bibitem{GanglEPD00}
M.~Gangl and H.~Ritsch, Eur.~Phys.~J.~D {\bf 8}, 29 (2000).

\bibitem{DohertyPRA00}
A.C.~Doherty {\it et. al}, Phys. Rev. A {\bf 63},
013401 (2000).

\bibitem{modefunctionref}
A.E. Siegman, {\it Lasers} (University science books, Mill Valley, CA 1986).

\bibitem{semipaper}
P.~Domokos, P.~Horak, and H.~Ritsch, J.~Phys.~B: At.~Mol.~Opt.~Phys. {\bf 34},
187 (2001).

\bibitem{FischerNJP}
T. Fischer, P. Maunz, T. Puppe, P.W.H. Pinkse, and G. Rempe, submitted to New
Journal of Physics. Here, the Langevin equations are solved for an ensemble of
atoms by defining an effective polarization, which has great similarities with
the effective mode approach.

\bibitem{GanglPRA00}
%{\it Cold atoms in a high-Q ring cavity}
M. Gangl and H. Ritsch, Phys. Rev. A {\bf 61}, 043405 (2000).

\end{references}
\end{document}